\documentclass[12pt]{iopart}

\usepackage{graphicx}
\usepackage{dcolumn}
\usepackage{bm}
\usepackage{amssymb}
\usepackage{braket}
\usepackage{hyperref}

\begin{document}
	
	\title[The initial stages of silver fluorination]{The initial stages of silver fluorination: a scanning tunneling microscopy investigation}
	
	\author{Jazm\'{i}n Arag\'{o}n S\'{a}nchez$^{1}$, Antonio Caporale$^{1}$, Ilya Degtev$^{2,3}$, Luciana Di Gaspare$^{4}$, Luca Persichetti$^{1}$, Maurizio Sansotera$^{5}$, Adri\'{a}n G\'{o}mez Pueyo$^{2,3}$, Monica De Seta$^{4}$, Jos\'{e} Lorenzana$^{2,3, *}$, Luca Camilli$^{1, \dag}$}
	\address{$^{1}$ Dipartimento di Fisica, Universit\`a di Roma ``Tor Vergata'', via della Ricerca Scientifica, 1-00133 Rome, Italy}
	\address{$^{2}$ ISC-CNR, Istituto dei Sistemi Complessi, via dei Taurini 19, 00185 Rome, Italy}
	\address{$^{3}$ Dipartimento di Fisica, Università di Roma ``La Sapienza'', 00185 Rome, Italy}
	\address{$^{4}$ Dipartimento di Scienze, Università Roma Tre, via Guglielmo Marconi 446, 00146 Rome, Italy}
	\address{$^{5}$ Dipartimento di Chimica, Materiali e Ingegneria Chimica “Giulio Natta”, Politecnico di Milano, Milan, Italy}
	
	\ead{$^{*}$jose.lorenzana@cnr.it,
		$^{\dag}$luca.camilli@uniroma2.it  (corresponding authors)}
	
	\vspace{10pt}
	\begin{indented}
		\item[]\today
	\end{indented}
	
	\begin{abstract}
		We use low-temperature scanning tunneling microscopy (LT-STM) to characterize the early stages of silver fluorination. On Ag(100), we observe only one adsorbate species, which shows a bias-dependent STM topography. Notably, at negative bias voltages, $V_{\rm B}<0$, the apparent shape can be described as a round protrusion surrounded by a moat-like depression (\textit{sombrero}). 
		As the voltage increases, the apparent shape changes, eventually evolving into a round depression. From the STM images, we determine the adsorption site to be the hollow position. 
		On Ag(110) we find adsorbates with three distinct STM topographies. 
		One type exhibits the same shape change with $V_{\rm B}$ as observed on Ag(100), that is, from a \textit{sombrero} shape to a round depression as the voltage changes from negative to positive values; the other two types are observed as round depressions regardless of $V_{\rm B}$. From the STM images, we find the three adsorbates to be sitting on the short-bridge, hollow and top position on the Ag(110) surface, with a relative abundance of 60\%, 35\% and 5\%. 
 	\end{abstract}
	
	\maketitle
	
	\section{\label{sec:Introduction}Introduction}
	Halogens are a group of chemical elements characterized by a remarkably high electronegativity, facilitating the formation of chemical compounds with a broad range of metals and the majority of nonmetals~\cite{Andryushechkin2018}. 
	Given their high sticking coefficient~\cite{Altman2001}, they are usually considered ideal elements for the study of the interaction between reactive gases and solid substrates~\cite{Zhu2016}. 
	In fact, the interaction of halogens, especially with metal surfaces, has been the subject of extensive investigation~\cite{Roman2014,Ignaczak1997,Tripkovic2009,Broekmann1999,Bechtold1981}, also in light of their technological potential, especially in the field of heterogeneous catalysis and microelectronics~\cite{Altman2001,Jones1988,Serafin1998}. 
	On many halogenated surfaces, a non-uniform arrangement of halogen adatoms has been observed at fractional monolayer coverages, often accompanied by substrate reconstruction~\cite{Komarov2023,Schott1994,Seliverstov2023}. 
	However, even though studies on the interaction between chlorine~\cite{Komarov2023,Baker2008}, bromine~\cite{Schott1994,Bange1987} or iodine~\cite{Bjork2013,Pacchioni1996} and transition metals, including Ag~\cite{Seliverstov2023,Kleinherbers1989,Koper1999}, are numerous, for the case of fluorine, only a very few experimental investigations have been reported so far~\cite{Bechtold1981,Schott1994,Schott1993,Qiu2000,Qiu2001,Barrena2021}, due to challenges associated with handling this highly corrosive element. Furthermore, most of the studied cases have been performed with diatomic gases whereas little is know on the reaction with monoatomic gases. 
	
	The case of silver fluorination is particularly interesting because, similar to oxygen with copper, fluorine can stabilize silver in the $d^9$ configuration~\cite{Grochala2001}. This has led to the emergence of a series of silver fluoride quantum materials with interesting magnetic properties~\cite{McLain2006,Miller2020,Sanchez-Movellan2021,Prosnikov2022,Wilkinson2023}. 
	In particular, AgF$_{2}$ is a correlated charge-transfer insulator, very similar to the parent compound of high-$T_{\rm c}$ cuprates~\cite{Gawraczynski2019,Bachar2022,Piombo2022,Yang2014}. 
	Moreover, it has been proposed that two-dimensional AgF$_{2}$ grown on specific substrates can host high-$T_{\rm c}$ superconductivity with $T_{\rm c}\sim 200$ K~\cite{Grzelak2020}. 
	Since a natural route for the synthesis of AgF$_{2}$ is silver fluorination~\cite{Ruff1934}, this provides a strong motivation to study this process from the early stages.
	
	An early experimental study~\cite{ODonnell1970} about the reaction of fluorine with silver samples was performed at high pressure of fluorine gas (50-600 Torr) and high temperatures (20-300 $^\circ$C), reporting  the synthesis of Ag$_{2}$F, AgF or AgF$_{2}$, depending on the experimental conditions. 
	Refs.~\cite{Schott1994,Schott1993} report scanning tunneling microscopy (STM) studies of atomically smooth halogen adlayers, including F, on Ag(111). Unfortunately, the growth method used in these studies did not allow to finely control the amount of halogens deposited on the samples, thus limiting the investigation only to fully halogenated surfaces. 
	Refs.~\cite{Qiu2000,Qiu2001} used X-ray photoemission spectroscopy (XPS) to characterize the fluorination of several transition metals with a fine control of the gas exposure. 
	However, the case of silver was not considered. 
	More recently, a study on the defluorination of C$_{60}$F$_{48}$ obtained silver fluorides as a byproduct \cite{Barre2002}. 
	Samples were studied by XPS and STM, but the surfaces were covered by the C$_{60}$
	molecules, which hampered a detailed characterization of the surface evolution as a function of fluorination.
	Therefore, as of today, an STM investigation of the first stages of fluorination of a metal substrate is still lacking.

	To fill this knowledge gap, here we report a low-temperature (LT)-STM characterization of the early stages of fluorination of low-index Ag surfaces.
    On the Ag(100) surface, we observe that F atoms adsorb on the hollow sites and show an STM apparent topography that is strongly dependent with tunneling conditions. On the other hand, on Ag(110), we find adsorbates sitting on hollow, short-bridge and top positions and each displaying a distinct STM topography.
    We discuss our findings in light of recent theoretical results~\cite{Pueyo2024}.

	\section{\label{sec: Experimental Methods}Methods}
	
	The studied samples consisted of Ag(100) and Ag(110) single crystals (purchased  from Surface Preparation Laboratory), which were cleaned by (1-2) keV Ar$^+$ sputtering cycles followed by annealing at 410 $^{\circ}$C in ultra-high vacuum (UHV).

	The fluorination was performed in a small ancillary chamber connected to the UHV system through a gate valve and constantly evacuated by a turbo molecular pump, reaching a base pressure of 5$\times$10$^{-9}$ Torr.
	A cylinder with a mixture of He/F$_2$ (nominal ratio of 9:1) was connected to the chamber via a leak valve. 
	Before loading the samples, the ancillary chamber was exposed to the He/F$_2$ mixture at room temperature and a pressure of 10$^{-5}$ Torr for $\sim$3 hours. This process passivates the walls, but more importantly, provides a fluorine-rich atmosphere to fluorinate the samples. 
	At the end of the passivation process, the pressure in the chamber was 1$-$2$\times$10$^{-8}$ Torr with monoatomic fluorine as the main gas component (see~\ref{sec:RGA}).  
	Thus, samples were placed into the ancillary chamber and exposed at room temperature for different amounts of time to this fluorine-rich atmosphere. Since, if not otherwise specified, all samples in this study were exposed to the same background pressure, when describing the different samples in the main text we only state their exposure time.
	After fluorination, the samples were moved to the STM chamber for characterization.

	STM experiments were performed using a commercial Infinity System (Scienta Omicron GmbH, Taunusstein, Germany) operating at $\sim9$ K and a base pressure better than $4\times10^{-11}$ Torr. 
	Electrochemically etched W tips were used for all STM measurements. 
	Data were collected in constant-current mode and calibrated using the lattice parameter values for Ag at the corresponding temperature as reported in Ref.~\cite{Hu2005}.
	Data were analyzed using Gwyddion software~\cite{NecasKlapetek2012}.
	
	\begin{figure}[tttt]
		\begin{center}
			\includegraphics[width=\columnwidth]{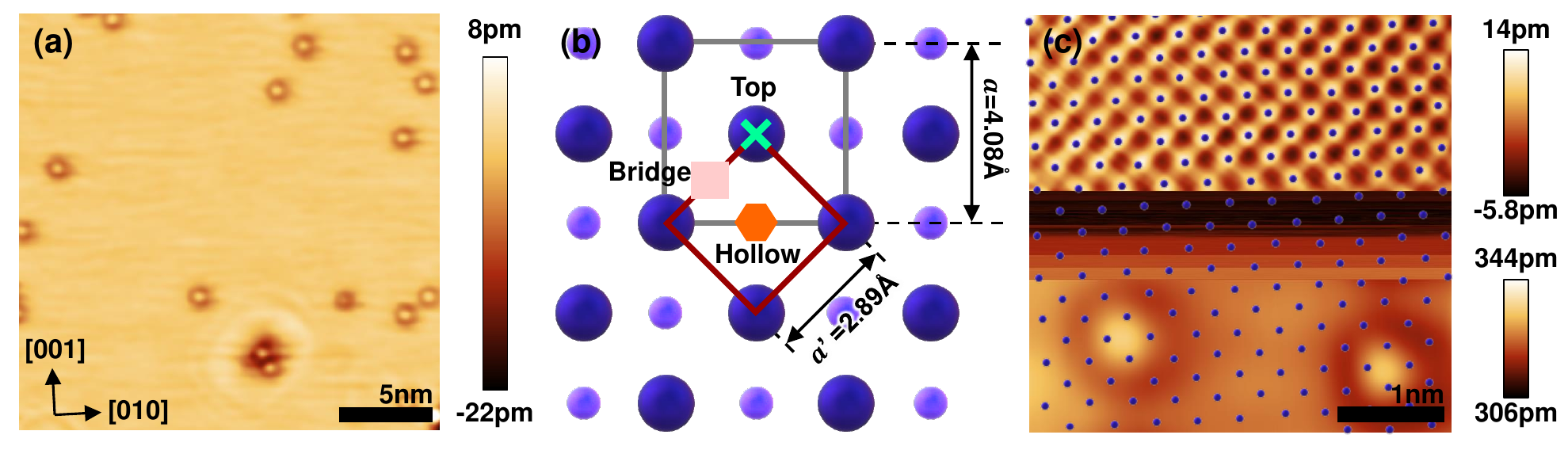}
		\end{center}
		\caption{(a) STM image of the Ag(100) crystal exposed 1.9 h to a fluorine-rich atmosphere, acquired at 9.3 K with bias voltage $V_{\rm B}=-0.5$ V and tunneling current 100 pA. 
			The crystallographic directions of the crystal surface are indicated by the black arrows in the bottom left corner of the image. These directions are the same for all images related to Ag(100). The large bright-contrast circle at the bottom of the image appears to be a defect of Ag. 
			(b) Schematic representation of the first two atomic layers of Ag(100). Surface atoms are depicted with large spheres, while atoms in the underlying layer are shown with small spheres. The red square indicates the unit cell for this surface. The lattice parameters, $a$ and $a'$, are based on experimental values obtained from X-ray diffraction at 9.3 K as reported in~\cite{Hu2005}. 
			The three highly symmetric positions on this surface are marked with different shapes: hollow (hexagon), bridge (square), and top (cross). 
			(c)  STM image showing both the F adatoms and atoms of the Ag(100) surface acquired at $V_{\rm B}=-0.5$ V. The tunneling current was set to 1 nA and 0.1 nA in the top and bottom half, respectively. The scale bar corresponds to 1 nm.} \label{Fig:STM_100_adatoms_large}
	\end{figure}
	
	\begin{figure}
		\begin{center}
			\includegraphics[width=\textwidth]{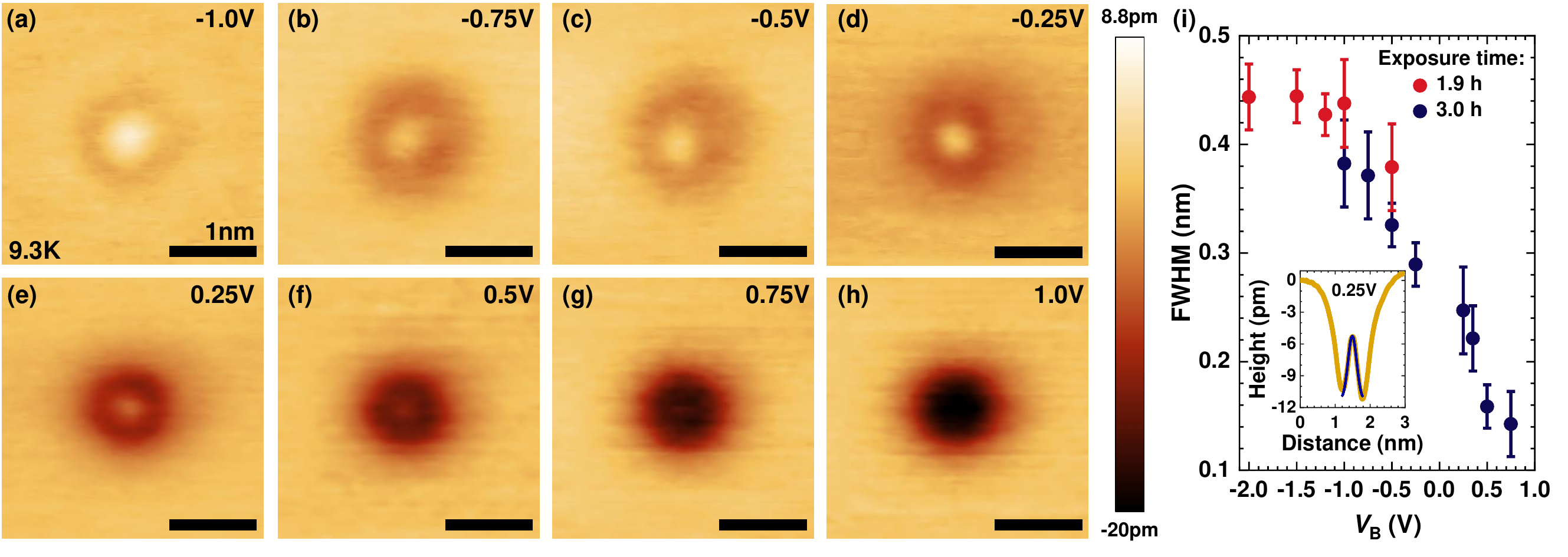}
		\end{center}
		\caption{(a)-(h) STM images of the same F adatom on Ag(100) as a function of the bias voltage. The images were acquired at 9.3 K for an exposure time of 3 h. The tunneling conditions were set at 750 pA and $-1<V_{\rm B}<1$ V. In all panels, the scale bar is 1 nm. 
			(i) Evolution of the FWHM Gaussian fit of the central protrusion with $V_{\rm B}$ for 1.9 (red dots), and 3 h (blue dots) exposure times.
			Inset: Example of Gaussian fit (blue line) of the central protrusion in the cross-sectional height profile obtained for the adatom shown in panel (e).} \label{Fig:STM_100}
	\end{figure}

	\section{Results}
	\label{subsec:ExperimentalResults}

	\subsection{\label{subsec:Ag100}Fluorination of Ag(100)}

	After exposing the Ag(100) crystal to the fluorine-rich atmosphere for 1.9 h, we observe the presence of isolated F adatoms on the substrate surface (Fig.~\ref{Fig:STM_100_adatoms_large}\,(a)). They appear as protrusions surrounded by a dark halo less than 1 nm wide forming a distinctive Mexican hat or \textit{sombrero} shape.

	The crystal structure of the Ag(100) surface is composed of atoms located in a square lattice with a lattice parameter $a'=a/\sqrt{2}$, with $a=4.08$ \AA{}~\cite{Hu2005}, see schematic of Fig.~\ref{Fig:STM_100_adatoms_large}\,(b). 
	On this surface, three possible highly-symmetric adsorption sites are present: hollow, bridge and top.
	In order to determine the adsorption site of F adatoms, 
	STM images as the one displayed in  Fig.~\ref{Fig:STM_100_adatoms_large}\,(c) were analyzed. 
	When collecting such an image, the bias was kept at $V_{\rm B}=-0.5$ V while the tunneling current was changed between a high setpoint (of the order of 1 nA), used for the top half without adatoms, and a low setpoint (of the order of 0.1 nA) for the bottom half where the adatoms were located. 
	This allows us to bring the STM tip very close to the surface in the top half of the image, thus obtaining atomic resolution of the Ag surface atoms (visible as bright spots), while minimizing interaction with the adatoms in the lower half. 
	The black lines at the center of the image are due to the change in tunneling current settings.
	The periodicity of the lattice was reproduced with blue dots in the bottom part of the image. Here, two fluorine adatoms are present, again with the \textit{sombrero} shape. 
	Clearly, the \textit{sombrero} apexes of both adatoms are located at the hollow sites of the Ag(100) surface. 
	This experimental finding agrees with previous theoretical studies about the adsorption of halogen atoms on Ag(100) finding the hollow site as the energetically most favorable one~\cite{Pueyo2024,Andryushechkin2009,Wang2002,Lamble1987,Fu2005}.  
	Incidentally, we note that a similar \textit{sombrero}-like topography was also observed for chalcogen atoms adsorbed on Ag(100)~\cite{Schintke2001,Spurgeon2019}.

	The apparent STM topography of isolated F adatoms on the surface strongly depends on the bias voltage, $V_{\rm B}$, and the tunneling current, $I_{\rm T}$.
	We first discuss the STM images collected with tunneling currents higher than 500 pA and later those taken with lower tunneling currents.  
	Panels (a)-(h) in Fig.~\ref{Fig:STM_100} show consecutive images of the same isolated F adatom measured with tunneling bias ranging from -1 to 1 V and a tunneling current of $I_{\rm T}=750$ pA. It is clear that the STM apparent shape of F adatoms  is a \textit{sombrero} when $V_{\rm B}<0.75$ V and  becomes a depression at 1 V. 
	The shape change is evident by looking at the STM pictures in Figs.~\ref{Fig:STM_100} (a)-(h). 
	At $V_{\rm B}=-1.0$ V, the F adatom presents the \textit{sombrero} shape with a maximum in the apex which protrudes out from the Ag surface (white region in the center of the topography in Fig.~\ref{Fig:STM_100} (a)).  
	It can be seen that, as $V_{\rm B}$ is increased, the height of the central protrusion of the \textit{sombrero} is continuously reduced as well as the the depth of the depression around it, until the \textit{sombrero} feature is lost, resulting in a smooth depression at 1.0 V, see Fig.~\ref{Fig:STM_100} (h).
    The same evolution of the appearance change is observed in the sequence of STM images presented in \ref{sec:Fadatom} for a F adatom measured at different bias voltages after a fluorination process of 1.9 h. 
	
	\begin{figure}
		\begin{center}
			\includegraphics[width=\columnwidth]{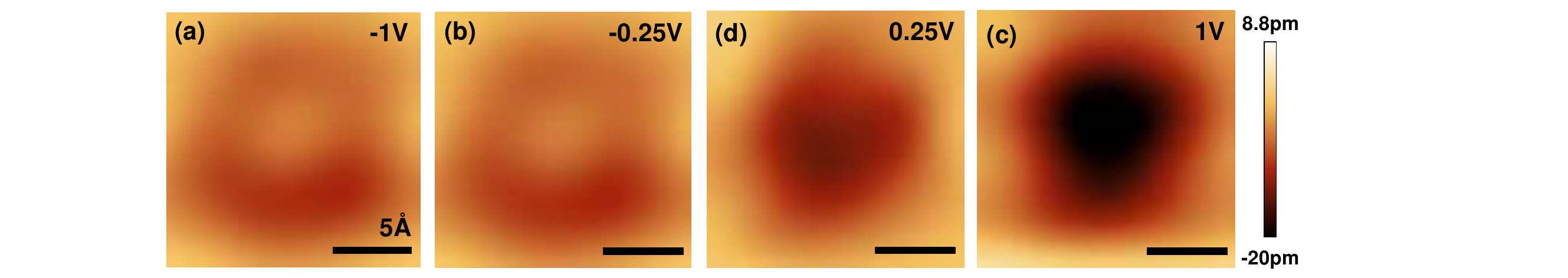}
		\end{center}
		\caption{STM images of an F adatom on Ag(100)  acquired at 9.4 K, 100 pA and $V_{\rm B}$= (a) -1, (b) -0.25, (c) 0.25 and (d) 1 V, after an exposure of 3 h. The scale bar corresponds to 5 \AA\, in each panel.} \label{Fig:SM_STM_100_Ad.pdf}
	\end{figure}

    The evolution of the topography can be compared with the theoretical results of Fig. 5 of Ref.~\cite{Pueyo2024}. The qualitative behavior is nearly the same.  A noticeable  difference is that  in  the theoretical computation the protrusion is always below the silver surface while in the experiment it overcomes the silver level for the lowest voltage.  Both results, however,  show clearly that there are two distinct contributions to the topography (the protrusion and the depression). 
	Theoretically the protrusion and the depression were associated to the oxidation of silver by fluorine. In particular, the protrusion was identified as the contribution of the fluorine $p-$orbitals which are filled and therefore below the Fermi level (negative voltage). Instead the depression was identified as due to the reduction of the electronic charge of silver (which becomes Ag$^+$). Such reduction reduces the screening of the nuclear charge and contracts the orbitals leading to a voltage independent depression~\cite{Pueyo2024}.

	In order to extract a quantitative indicator for this shape transition, the central protrusion in the cross-sectional height profiles was fitted to a Gaussian function, as shown by a blue line in the inset of Fig.~\ref{Fig:STM_100}\,(i).
	The fitted Gaussian full-width-at-half-maximum (FWHM) as a function of $V_{\rm B}$ is presented for the two different exposure times in the main panel of this figure.
	The FWHM shows two regimes in the analyzed $V_{\rm B}$ range.   
	For $-2 < V_{\rm B} < -1$ V, the FWHM remains approximately constant at a value of $(0.45\pm 0.03)$ nm.
	This result is quantitatively similar to the one reported for S atoms adsorbed on Ag(100) crystals, where a FWHM value of $(0.38\pm0.04)$ nm was measured in the same $V_{\rm B}$ range~\cite{Spurgeon2019}. 
	Conversely, the FWHM value monotonically decreases from 0.4 to 0.14 nm as $V_{\rm B}$ is increased from -1 to 0.75 V.
	In the S/Ag(100) system, a similar result was explained in terms of the through-surface and through-adsorbate conductances, which define the height of the silver surface and the height at the point directly above the adsorbate, respectively.  When $V_{\rm B}$ increases, the through-surface conductance increases much more rapidly than the through adsorbate-conductance, so that the adsorbate height drops below that of the silver surface~\cite{Spurgeon2019}. Ref.~\cite{Pueyo2024} provides a more detailed explanation in terms of the density of states of the F $p$- and Ag $s$-orbitals and their respective extensions.  
	
	At low tunneling currents, the STM images show the same qualitative behavior but the threshold voltage for the disappearance of the \textit{sombrero} feature decreases. 
	Figure~\ref{Fig:SM_STM_100_Ad.pdf} shows a sequence of topographies measured at 100\,pA.
	The appearance of an isolated F adatom on Ag(100) surface changes from \textit{sombrero} to depression when $V_{\rm B}$ is increased from negative to positive values. 
	This behavior contrasts with the one observed in Fig.~\ref{Fig:STM_100}, where the F adatom is detected as a \textit{sombrero} up to V$_{\rm B}=$ 0.75 V.

	\subsection{Fluorination of Ag(110)} 
	
		\begin{figure}
		\centering
		\includegraphics[width=\columnwidth]{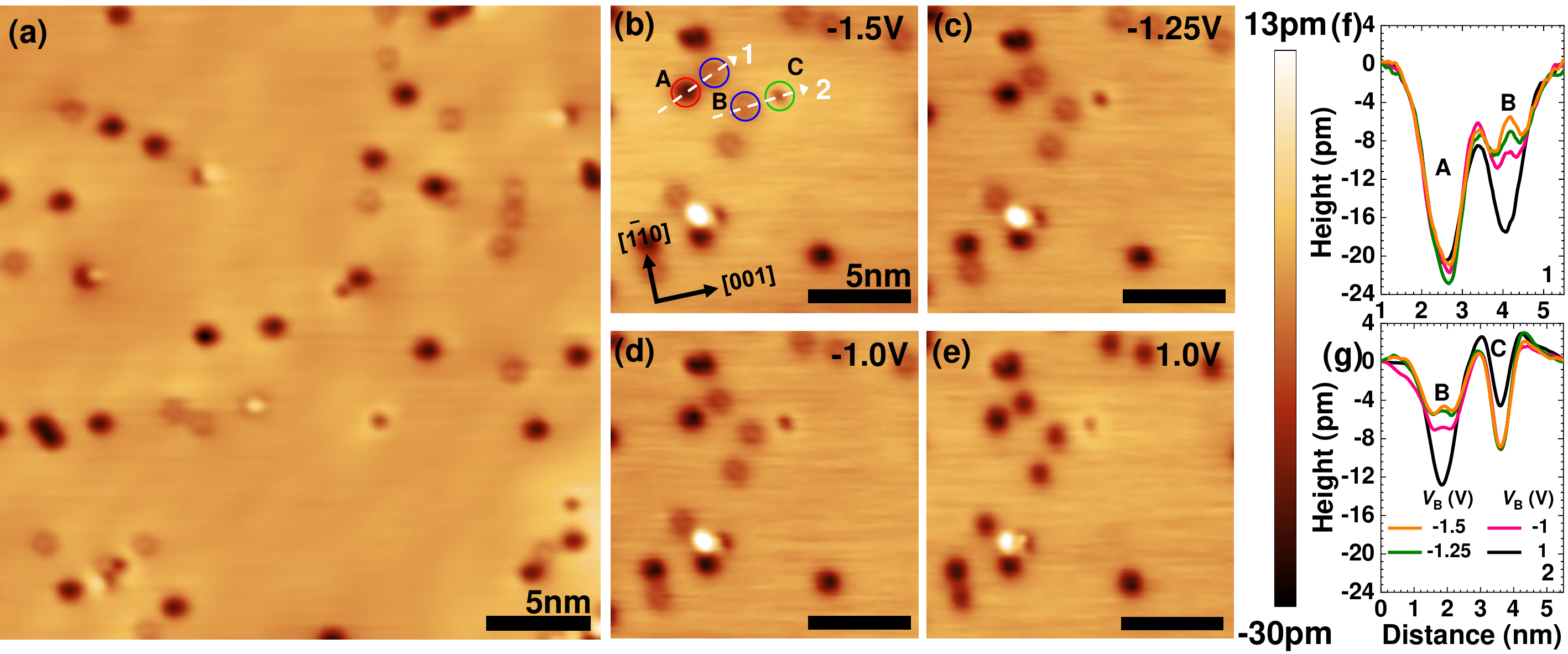}
		\caption{(a)-(e) STM image of the Ag(110) surface acquired at 9.3 K after an exposure to a fluorine-rich atmosphere for 45 minutes. The tunneling conditions were set to $V_{\rm B}=-0.5$ V and 80 pA in panel (a). The tunneling current was set to 35 pA and $V_{\rm B}$ varies from -1.5 to 1 V. 
		The scale bar corresponds to 5 nm in all panels. 
		Blue, red, and green circles in panel (b) highlight the different features observed in STM images acquired at negative bias voltages.
		(f)-(g) Height profiles for different $V_{\rm B}$ values acquired along the white dashed arrows 1 and 2 indicated in panel (b), respectively. Letters A, B, and C in these panels indicate the different species indicated in panel (b).}
		\label{fig:STM_110_adatom}
	\end{figure}

	After exposing the Ag(110) crystal to the fluorine-rich atmosphere for 45 minutes, the surface results in being covered by adsorbates exhibiting different apparent topographies (Fig.~\ref{fig:STM_110_adatom}\,(a)). In Figs.~\ref{fig:STM_110_adatom}\,(b)-(e), the same area was measured using a tunneling current of 50 pA and voltage bias in the range $-1.5\leq V_{\rm B}\leq 1$ V. Specifically, three distinct types of apparent topographies are identified and marked in Fig.~\ref{fig:STM_110_adatom}\,(b) as type-A (red circle), type-B (blue circles), and type-C (green circle). The bright-contrast feature in the bottom left part of the image is not considered as it appears rarely, thus it is not statistically relevant.
	Height profiles for these adsorbates, recorded at different bias voltages along the lines indicated by the white arrows in panel (b), are shown in panels (f) and (g). 

\begin{figure}
		\begin{center}
			\includegraphics[width=\columnwidth]{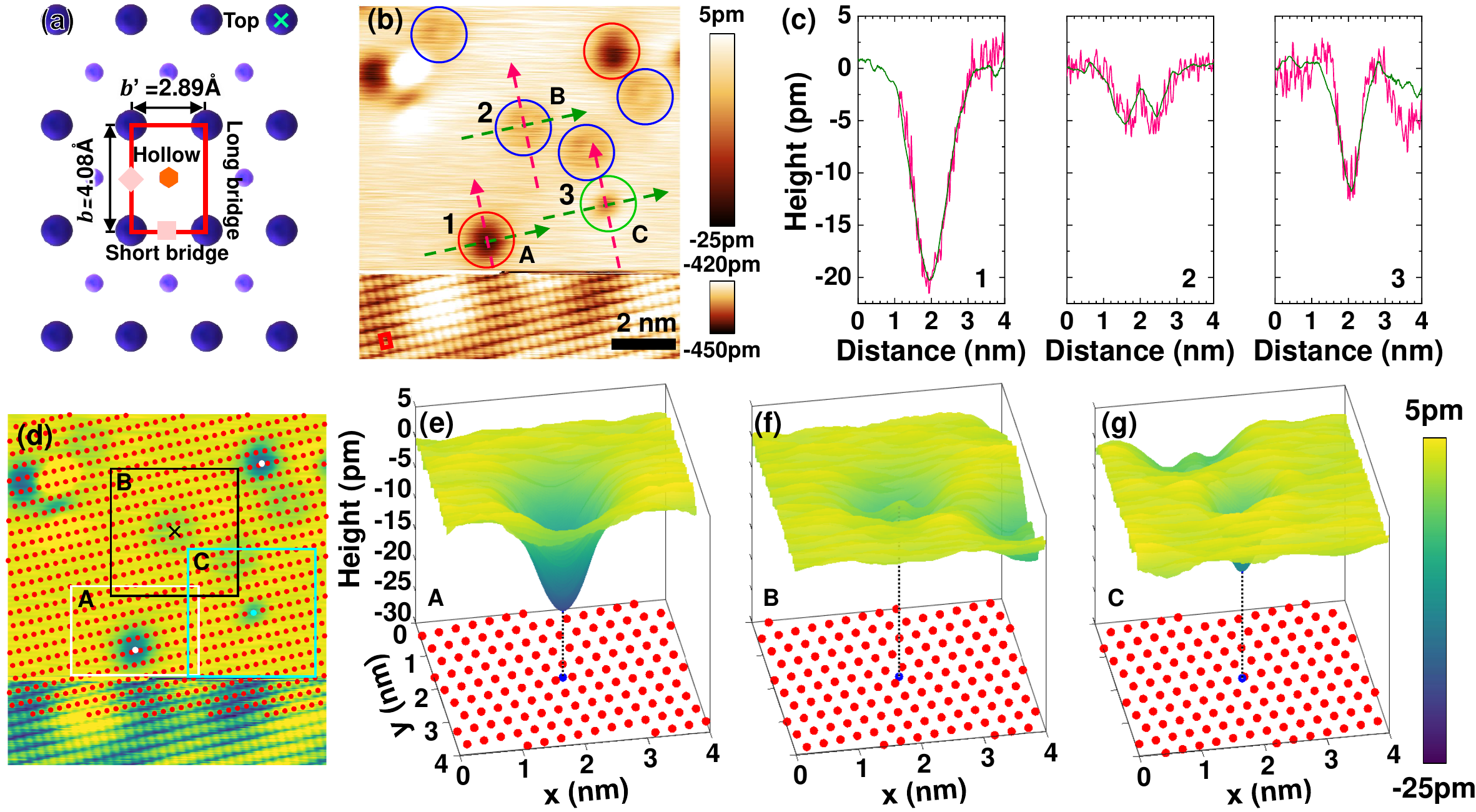}	
		\end{center}
		\caption{(a) Schematic representation of the first two surface layers of Ag(110). Large spheres represent surface atoms, while small spheres depict atoms in the subsurface layer. The red rectangle outlines the unit cell of the surface. Lattice parameters, denoted as $b$ and $b'$, were determined from XRD patterns, reported in~\cite{Hu2005}. The four primary high-symmetry adsorption sites are indicated: hollow (hexagon), short-bridge (square), long-bridge (rhombus) and top (cross).
        (b) STM image showing some adsorbates in the top part and atomic resolution of the Ag(110) surface in the bottom part, acquired at $V_{\rm B}=-0.01$ V.
		The tunneling current was set to 0.1 and 1.0 nA in the top and bottom part, respectively. 
		The green and pink dashed arrows indicate the directions along which the height profiles shown in panel (c) were taken. (d) Flattened topography of the same region shown in panel (b), optimized for better visualization.
		Red dots mark the positions of surface silver atoms, while white circle, black cross and cyan circle indicate the centers of a type-A, type-B and type-C adsorbates, respectively.
		The areas marked by the colored frames A, B and C are shown as three-dimensional renderings in the panels (e), (f) and (g), respectively.
		In each plot, a blue dot at the bottom highlights the position of the corresponding adsorbate.} 
		\label{Fig:STM_110}
	\end{figure}

    It is worth mentioning that the Ag atoms on the (110) surface form a rectangular unit cell, as highlighted by a red frame in the schematic shown in Fig.~\ref{Fig:STM_110}\,(a).
	The lateral dimensions of the unit cell are $b=4.08$ \AA{} and $b'=2.89$ \AA{}. 
	This geometry presents four possible high-symmetry adsorption sites: hollow, top, short bridge, and long bridge. 
	According to DFT calculations, the hollow and both bridges sites (short and long) are the most energetically favorable for F adatoms, with comparable adsorption energies, while there is a significant energy penalty for the top position ~\cite{Pueyo2024,Wang2001}.
	
	Type-A adsorbates appear as depressions, with their apparent shape largely independent of bias over the presented range ($-1.5 \leq V_{\rm B} \leq 1$ V), (Fig.~\ref{fig:STM_110_adatom}\,(f), left feature). 
	Type-B adsorbates (right and left features in Figs.~\ref{fig:STM_110_adatom}\,(f) and (g), respectively) exhibit a \textit{sombrero} shape at negative bias, that is lost at $V_{\rm B}=1.0$ V.
	Type-C adsorbates are detected as depressions surrounded by a brighter halo, resulting in a distinctive volcano-like profile, with a central minimum and two lateral local maxima (right feature in Fig.~\ref{fig:STM_110_adatom}\,(g)).
	For $V_{\rm B} < 0$, the depth of the depression and that of the surrounding region remain unchanged, but both of them increase at $V_{\rm B} = 1.0$ V. At $V_{\rm B} > 0$, type-A adsorbates have the deepest height (-24 pm), followed by the type-B (-16 pm), and type-C (-4 pm), as shown in Figs.~\ref{fig:STM_110_adatom}\,(f) and (g).
	However, for $V_{\rm B} < 0$, type-C adsorbates appear deeper than type-B due to the central protrusion of the \textit{sombrero}-shaped type-B feature.

	STM images, such as the one shown in Fig.~\ref{Fig:STM_110}, were analyzed to pinpoint the adsorption sites of the adsorbates.
	In the top part of the topography, the three types of adsorbates are identified and marked: A (red circles), B (blue circles) and C (green circle). 
	Since this image was acquired at $V_{\rm B} < 0$, type-B adsorbates appear with a \textit{sombrero} shape, while type-A and type-C adsorbates are observed as depressions with different depths. 
	The deepest point for type-A and type-C adsorbates and the apex in the center of type-B adsorbates were determined from the left, right and center height profiles in Fig.~\ref{Fig:STM_110}\,(c), respectively, and taken as the position of the adsorbates.

	To determine the position of the Ag surface atoms in the top part of the image, the lattice observed in the bottom part was reproduced and marked with red dots in Fig.~\ref{Fig:STM_110}\,(d). 
	In this panel, the positions of the centers of type-A, type-B  and type-C  adsorbates, as extracted from the line-profiles in panel (c), are visible as white dots, a black cross and a cyan dot, respectively.
	This arrangement is better visualized in the three-dimensional representations displayed in Figs.~\ref{Fig:STM_110}\,(e)-(g), and corresponding to the framed areas reported in panel (d). Here, the blue dot marks the position of each adsorbate, while the dashed line serves as a guide to the eye, projecting this position onto the Ag(110) surface atomic plane (red dots). 
     This shows that type-A adsorbates are located at the short-bridge sites on the Ag(110) surface, type-B at the hollow sites, whereas type-C at the top sites. A statistical analysis performed on several STM images reveals that the type-A adsorbates are the most abundant (60\%), followed by type-B (35\%). Type-C adsorbate is the least frequent, with only 5\% of relative abundance.
 
	The coexistence of the three observed species could be explained by assuming that the F atoms populate different adsorption sites on the Ag(110) surface. This is in fact compatible with DFT calculations predicting that the short-bridge, long-bridge, and hollow adsorption positions have adsorption energies which differ by less than 150 meV~\cite{Pueyo2024,Wang2001}. However, in our experiments, we do not find adsorbates occupying long-bridge sites, whereas we observe a small minority of them occupying the top sites. Because DFT calculations~\cite{Pueyo2024,Wang2001} find a large energy penalty for F adatoms on top positions, other possibilities to explain the presence of type-C adsorbates need to be considered. It is indeed worth recalling that HF is also present in the fluorine-rich atmosphere to which the Ag crystals were exposed during the fluorination treatment (\ref{sec:RGA}). Incidentally, the HF partial pressure, which is almost one order of magnitude smaller than the F partial pressure, might explain the low abundance of type-C adsorbates. However, other possibilities, for example substitutional F (that is, F atom replacing a Ag surface atom) cannot be ruled out and therefore  we believe that additional DFT calculations are needed to resolve this issue.

	In our experiments, we estimate a coverage up to $\theta=0.006$ ML of F adatoms on both surfaces (see \ref{sec:Sticking} for an example on the Ag(110) surface). From the measured partial pressure and exposure times we estimate a sticking coefficient $s_0\sim10^{-4}$ or even smaller, depending on the surface. This value is subject to uncertainties as, for example, the initial pressure estimated in 10$^{-8}$ Torr can decrease down to 20\% during the exposure. However, such variations do not change the order of magnitude, which is sufficient for the discussion.  This small value is surprising, as halogens are generally expected to exhibit large sticking coefficients, typically around one, as mentioned in the introduction. One possible explanation for this discrepancy is that our study used atomic fluorine, whereas most previous studies investigated diatomic halogen gases. At first glance, this makes the result even more puzzling. Thermodynamically, atomic fluorine should be more reactive: according to DFT computations~\cite{Pueyo2024}, the adsorption energy per atom for atomic fluorine is 4.2 eV, while the bond cleavage cost for F$_{2}$ is 1.6 eV. On the other hand, the resulting excess energy must be dissipated for the F atoms to be effectively trapped, and this energy dissipation is more efficient in the case of diatomic molecules since one atom can be adsorbed while the other atom can be ejected carrying the excess energy. This process known as atom abstraction is known to occur in the case of silicon exposed to diatomic fluorine~\cite{Li1995}. Further theoretical and experimental investigations are needed to resolve this puzzle.

	\section{\label{sec:Conclusions}Summary and Conclusions}
	In this work, we have studied the early stages of silver fluorination on the (100) and (110) surfaces. Silver crystals were exposed to a fluorine-rich atmosphere at room temperature for varying periods of time ($\leq 3$ h). Subsequently, the samples were characterized at low temperature using STM.

	On both surfaces, a low density of individual adsorbates were observed.
	On the Ag(100) surface, only one type of adsorbate is observed, with an apparent topography that strongly depends on the tunneling conditions. For $V_{\rm B}<0$, the adsorbates appear as \textit{sombreros}, meaning round protrusions surrounded by a circular depression. As $V_{\rm B}$ increases, the height of the \textit{sombrero's} apex tends to decrease until it disappears, leaving only a depression in the topography. 
	The \textit{sombrero}-to-depression transition occurs at a positive bias voltage value that depends on the applied tunneling current.
	Analysis of atomically resolved STM images allowed us to determine that the \textit{sombrero's} apex is located at the hollow site of the Ag(100) surface. 
	This result is consistent with DFT studies~\cite{Pueyo2024,Wang2001}, which predict the hollow adsorption site on this surface to be by far the most energetically favorable one for F adatoms.

	On the Ag(110) surface, adsorbates are detected with three different apparent topographies. Type-A adsorbates appear as depressions for the whole range of $V_{\rm B}$ studied, with a weak dependence on the bias voltage; type-B adsorbates exhibit a \textit{sombrero}-to-depression transition as a function of $V_{\rm B}$, similar to what is observed for F/Ag(100); finally, type-C ones appear as round depressions surrounded by a bright halo, with its appearance only weakly affected by $V_{\rm B}$. 
    From our STM images, we determine that type-A adsorbates are located at the short-bridge site, type-B at the hollow site, and type-C at the top site of the Ag(110) surface.  
	The relative abundance of these three species is found to be 60\%  for type-A, 35\% for type-B and only 5\% for type-C. Leveraging on previous theoretical considerations~\cite{Pueyo2024,Wang2001} that found a large energy penalty for F adatoms sitting on top sites, type-A and type-B adsorbates are assigned to F adatoms whereas the attribution of type-C is more challenging and requires further theoretical investigations.

	It is worth mentioning that this work reports the first STM investigation of the early stages of fluorination of a transition metal, specifically Ag. In this context, and considering the recent interest in AgF$_{2}$ as possible high-$T_{\rm c}$ superconductor~\cite{Grzelak2020}, the rich dataset shown here is a valuable addition to current knowledge and will bolster additional studies on this topic.

	\ack
	We thank Y. Fasano and R. Larciprete for their insightful comments and discussions. 
	Work supported by the Italian Ministry of University and Research through the project 
	Quantum Transition-metal FLUOrides (QT-FLUO)
	PRIN 20207ZXT4Z and CNR-CONICET NMES project.
    This work was supported by the Italian Ministry for Universities and Research (MUR) under the Grant of Excellence Departments (Article 1, Paragraphs 314–337, Law 232/2016) to the Department of Science, Roma Tre University.

	\appendix
	\section{\label{sec:RGA}Residual gas analysis of the fluorination chamber}
	
	 	\begin{figure}[hhhh]
		\begin{center}
			\includegraphics[width=\columnwidth]{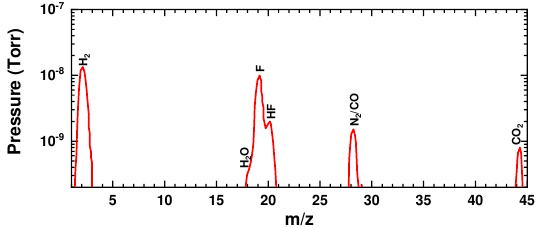}
		\end{center}
		\caption{Residual gas analysis measured up to mass-to-charge ratio $m/z=45$  in the ancillary chamber right after a passivation process.} \label{Fig:RGA}
		\end{figure}
	
	In this study, we present the characterization of F atoms adsorbed on Ag(100) and Ag(110) using low-temperature scanning tunneling microscopy (LT-STM). To perform the fluorination process, the Ag(100) and Ag(110) samples were placed in an ancillary chamber where they were exposed to a fluorine-rich atmosphere. The chamber is equipped with a residual gas analyzer (RGA) (RGA100 from Stanford Research Systems).
	As reported in Fig.\,\ref{Fig:RGA}, in addition to the species typically found in UHV systems—namely H$_2$, N$_2$/CO, CO$_2$, and small traces of H$_2$O, with mass-to-charge ratios of 2, 28, 44, and 18, respectively—F and HF with mass-to-charge ratios of 19 and 20, respectively, can also be observed.
	If one excludes H$_2$, monoatomic F (19) is the most abundant residual gas
	in the ancillary chamber, with its partial pressure being at least almost one
	order of magnitude higher than the partial pressure of the other residual
	gases.

    \section{\label{sec:Fadatom}STM images of isolated F adatom on Ag(100) as a function of the bias voltage}

    \begin{figure}[hhhh]
        \centering
        \includegraphics[width=\linewidth]{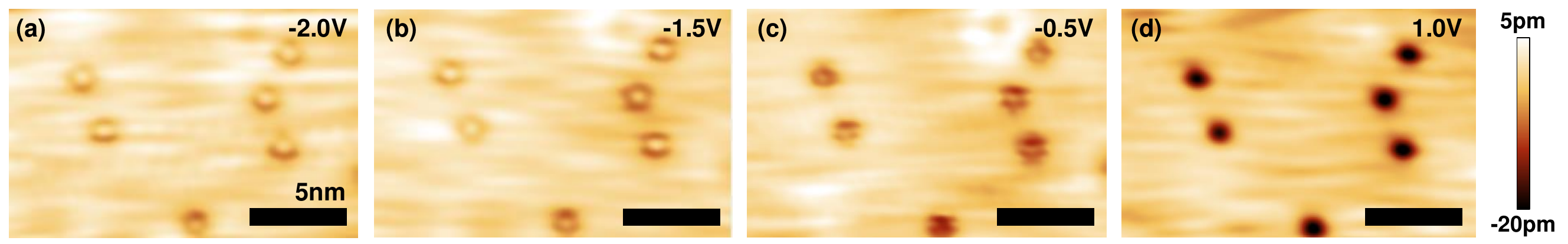}
        \caption{LT-STM images of single F adatoms on Ag(100) surface after an exposure time of 1.9 h (200 pA, $-2.0 \leq V_{\rm B}\leq 1.0$ V). The scale bar corresponds to 5 nm in each panel.}
        \label{fig:STM_100_adatoms_appex}
    \end{figure}

    Typical STM images of F adatoms on Ag(100) are shown in Fig.\,\ref{fig:STM_100_adatoms_appex}. These topographies were recorded in the same area with bias voltages ranging from -2.0 to 1.0 V and a tunneling current of $I_{\rm T} = 200$ pA. For $V_{\rm B}<0$, F adatoms appear as \textit{sombreros}, following a transition from \textit{sombrero} to depression as observed after a 3 h exposure, as shown in Figs.\,\ref{Fig:STM_100} (a)-(h) in \ref{subsec:Ag100}.

    Cross-sectional height profiles of each F adatom were fitted to a Gaussian function to obtain the full-width-at-half-maximum (FWHM) as a function of $V_{\rm B}$. The resulting data are plotted in Fig.\,\ref{Fig:STM_100} (i).

     \section{\label{sec:Sticking}Estimation of F adatoms coverage on the Ag(110) surface}

     \begin{figure}[hhhh]
         \centering
         \includegraphics[width=\columnwidth]{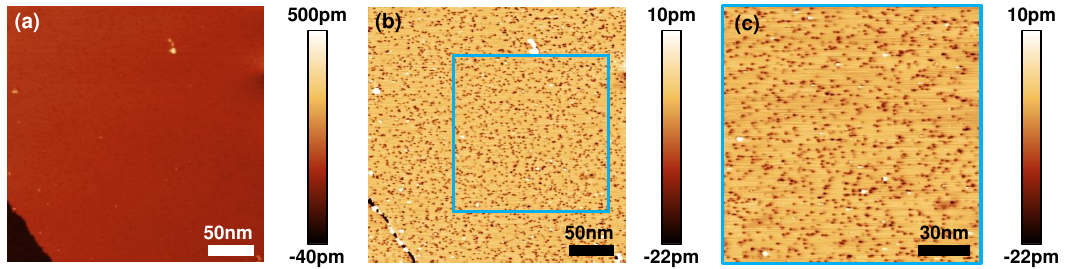}
         \caption{(a) STM image of fluorinated Ag(110) surface acquired after an exposition of 45 min to a fluorine-rich atmosphere. The tunneling conditions were set to 60 pA and –1.5 V. (b) Flattened STM image presented in (a). The step is observed as the dark line in the bottom left corner. The framed area is shown in (c).}
         \label{fig:Big_110}
     \end{figure}

    In order to estimate the coverage, $\theta$, on the studied surfaces, we analyzed STM images over large scanned areas. Figure~\ref{fig:Big_110} presents an example of the fluorinated Ag(110) surface after 45 minutes of exposure. Even over this large area, no formation of clusters or islands is observed. Instead, the adatoms tend to remain isolated, as seen in the zoomed-in view in panel (c) of this figure. For this fluorination experiment, we estimate a density of (0.053$\pm0.002$) adatoms$/$nm$^{-2}$. Considering the unit cell area of Ag(110) surface, we obtain a coverage $\theta=0.0062\pm0.0001$.

\end{document}